\documentclass[amsmath,amssymb,reprint,twocolumn,superscriptaddress, longbibliography, nobibnotes, prb]{revtex4-1}

\usepackage[utf8]{inputenc}
\usepackage[T1]{fontenc}
\usepackage[english]{babel}
\usepackage{changes}   
\usepackage[colorlinks=true, citecolor=blue, urlcolor=blue, linkcolor=blue]{hyperref} 

\usepackage{amsmath,amssymb,amsfonts}
\usepackage{amstext, mathrsfs, textcomp}
\usepackage{mathptmx}
\usepackage{bigints}  
\usepackage{nicefrac}

\usepackage{multirow}
\usepackage{dcolumn}
\usepackage{bm}

\usepackage{subfigure}
\usepackage{graphicx}
\usepackage{xcolor}
\usepackage[export]{adjustbox}

\graphicspath{{Figures/}{graphical_TOC/}}

\newcommand{\TITLE}{Data-driven Azimuthal RHEED construction for in-situ crystal growth characterization}

\begin{document}
\title{\TITLE}
\author{\firstname{Abdourahman} \surname{Khaireh-Walieh}}
\email[e-mail~: ]{akhairehwa@laas.fr}
\affiliation{LAAS-CNRS, Universit\'e de Toulouse, 31400 Toulouse, France}

\author{\firstname{Alexandre} \surname{Arnoult}}
\affiliation{LAAS-CNRS, Universit\'e de Toulouse, 31400 Toulouse, France}

\author{\firstname{Sébastien} \surname{Plissard}}
\affiliation{LAAS-CNRS, Universit\'e de Toulouse, 31400 Toulouse, France}

\author{\firstname{Peter} \surname{R. Wiecha}}
\email[e-mail~: ]{pwiecha@laas.fr}
\affiliation{LAAS-CNRS, Universit\'e de Toulouse, 31400 Toulouse, France}

\makeatletter
\let\doctitle\@title  
\makeatother

\begin{abstract}
Reflection High-Energy Electron Diffraction (RHEED) is a powerful tool to probe the surface reconstruction during MBE growth. However, raw RHEED patterns are difficult to interpret, especially when the wafer is rotating. A more accessible representation of the information is therefore the so-called Azimuthal RHEED (ARHEED), an angularly resolved plot of the electron diffraction pattern during a full wafer rotation. However, ARHEED requires precise information about the rotation angle as well as of the position of the specular spot of the electron beam. We present a Deep Learning technique to automatically construct the Azimuthal RHEED from bare RHEED images, requiring no further measurement equipment. We use two artificial neural networks: an image segmentation model to track the center of the specular spot and a regression model to determine the orientation of the crystal with respect to the incident electron beam of the RHEED system. Our technique enables accurate, and potentially real-time ARHEED construction on any growth chamber equipped with a RHEED system.
\end{abstract}

\maketitle
\section{Introduction}

Molecular Beam Epitaxy (MBE) is a growth method allowing for synthesis of crystals with atomic precision and fabrication of advanced heterostructures \cite{jager_mbe_2011, steinle_monolithic_2001}. Moreover, in-situ tools such as Reflection High Energy Electron Diffraction (RHEED) \cite{ino_new_1977, braun_applied_1999, horio_new_1996, ichimiya_reflection_2004}, reflectometry \cite{corfdir_monitoring_2018, bardinal_precision_1995}, curvature \cite{arnoult_magnification_2021} or bandgap measurements \cite{johnson_situ_1998} can be used for real-time control of the growth processes, which is unmatched by other growth methods that often struggle to accurately determine growth conditions in-situ.
In MBE, source elements (e.g. Gallium or Arsenic) are individually heated to a temperature where evaporation/sublimation takes place, which creates a beam of atoms or molecules. These beams are directed towards a crystalline, heated substrate where they are adsorbed, diffuse on the surface and finally chemically bond. RHEED patterns provide information about the crystalline surface with atomic resolution and are highly sensitive to several key MBE parameters such as the growth rate, the temperature, the crystal structure, the lattice parameter, the strain, etc.  \cite{jo_real-time_2018, ohtake_strain_2020, hartmann_reflection_1998}
However, raw RHEED patterns are difficult to interpret directly and therefore are typically preprocessed. An interesting way to present RHEED patterns is the so-called Azimuthal Reflection High-Energy Electron Diffraction  (ARHEED), in which slices through the specular spot of RHEED patterns are drawn in a polar plot as a function of the azimuthal angle (the rotation angle of the substrate) \cite{ingle_structural_2010}. ARHEED offers a relatively straightforward interpretability and allows to detect for instance periodicities or epitaxial crystal orientations.

\begin{figure*}[t]
    \centering
    \includegraphics[width=.99\linewidth]{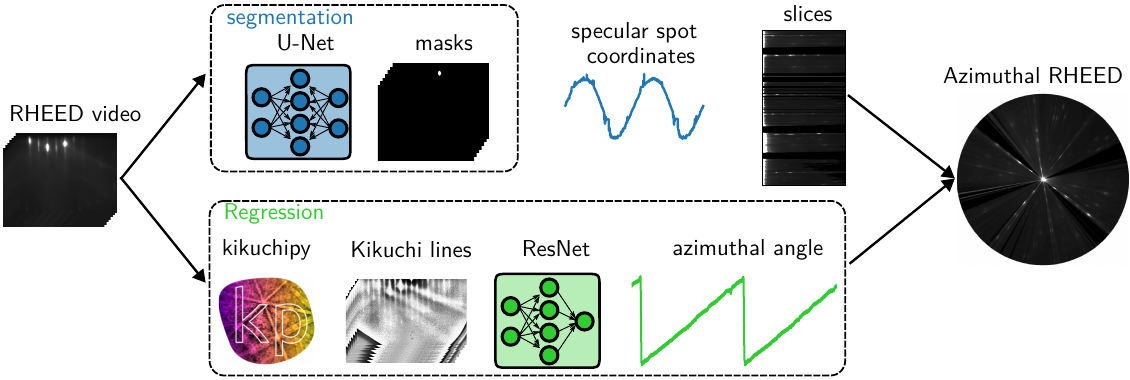}
    \caption{\small Azimuthal RHEED construction. Short videos of RHEED patterns go through two parallel paths. The first path, on the top, consists of segmentation using U-Net to produce binary masks of the specular spot in or order to retrieve the coordinates of the gravity center. The specular spot positions are used to crop thin slices from the video frames. The second path, on  the bottom, preprocesses the images with kikuchipy \cite{anes_processing_2020} with the goal of highlighting the Kikuchi lines. The data is then fed to a ResNet model for azimuthal angle regression. The Azimuthal RHEED is obtained by polar plotting the slices through the specular spot in function of the corresponding azimuthal angles.}
    \label{fig:overview_figure}
\end{figure*}

During MBE, it is essential to grow layers with substrate rotation to ensure uniformity \cite{braun_reflection_1998}. Therefore, it is important to develop methods that allow access to the reciprocal lattice of rotating substrates \cite{braun_reflection_1998}. 
Xiang et al. have showed in \cite{xiang_reflection_2016} that it is possible to obtain the entire reciprocal space structure of a 2D material by rotating the sample around the surface normal and measuring the RHEED patterns as a function of the azimuthal angle which makes ARHEED extremely interesting for surface characterization. 
A frequent challenge is the fact that a typical MBE chamber does not have a device for azimuthal angle measurement; and when it does, a typical difficulty is associated with the fact that in the ultra-high vacuum MBE chambers, the rotational stage is coupled magnetically and the rotation is often not perfectly smooth. 
This makes it difficult to determine the crystal angle with good accuracy. Determining the crystal azimuthal angle from the RHEED pattern itself is therefore very interesting for high-accuracy ARHEED. 
The Kikuchi lines, illustrated in Figure \ref{fig:kikuchi_lines_and_model_architecture}a-c with red lines, contain information on the sample orientation. Considering standard (001) oriented substrates, they are (for samples with a cubic crystal lattice) 4-fold symmetric and thus strictly contain information only for 0-90 degrees. 
Using several consecutive RHEED images instead of single ones to feed the neural network model would include information about the direction of the Kikuchi lines rotation and thus would allow to identify 0-180 degrees. Finally, the RHEED screen is usually not perfectly centered, such that each quadrant of the rotation shows a different section of the Kikuchi lines pattern, thus it should even be possible to distinguish between a full 0-360 degrees rotation angle range. Kikuchi lines originate from the electron scattering in the bulk of the crystal, hence do not (or only very weakly) depend on the growth of the topmost monolayer. Determination of the rotation angle should thus be possible in a robust way. 
Additionally, also the surface signal (diffraction spots and lines) contains information about the crystal angle, which can be used by a neural network as already demonstrated in earlier works \cite{khaireh-walieh_monitoring_2023, kwoen_classification_2020, kwoen_multiclass_2022, pricePredictingAcceleratingNanomaterial2024}. 
In summary, two elements are necessary in order to construct the ARHEED from raw RHEED patterns: the coordinates of the specular spot across the RHEED patterns and the rotational angle. Therefore, our approach in this article aims to construct the ARHEED from raw RHEED images by detecting the specular spot with segmentation using U-Net and determining the azimuthal angle via regression using ResNet. An overview of the proposed workflow is depicted in Figure \ref{fig:overview_figure}.

\section{Results and discussion}
\label{sec:results} 

\subsection{Specular spot tracking across RHEED patterns} \label{sec:Specular spot tracking across RHEED patterns}

In order to crop thin slices from RHEED images through the specular spot, its position must be tracked across the RHEED patterns. Generally, the tracking process involves two algorithms, one for object detection \cite{kaur_comprehensive_2023, roy_fast_2022, liu_survey_2021} and one for object tracking \cite{chen_visual_2022, zhang_recent_2021}. Object detection algorithms firstly identify and localize objects in images followed by object tracking algorithms that link the detected objects across video frames to maintain their identity and trajectory. In case of object disappearance, the process of object detection is relaunched.

In the case of RHEED, the specular spot is lost from view at least four times per rotation as a consequence of the sample holder hooks, blocking the electron beam. Furthermore, the specular beam disappears occasionally due to fluctuations in light conditions and noise. This would require the detection algorithm to be restarted each time a loss occurred. Another choice is to process each image independently performing a segmentation task to divide the image into meaningful regions \cite{yu_techniques_2023}. Unlike object detection, which identifies and localizes objects using bounding boxes, the segmentation provides a detailed pixel-level identification. In our study, the specular spot pixels are referred to as "class one" whereas the remaining area is referred to as "class zero" leading to a binary mask from each RHEED image independently. This is achieved by implementing a lightweight U-Net architecture of 84k parameters.

\subsubsection{Dataset preprocessing}

\begin{figure}[h]
    \centering
    \includegraphics[width=\linewidth]{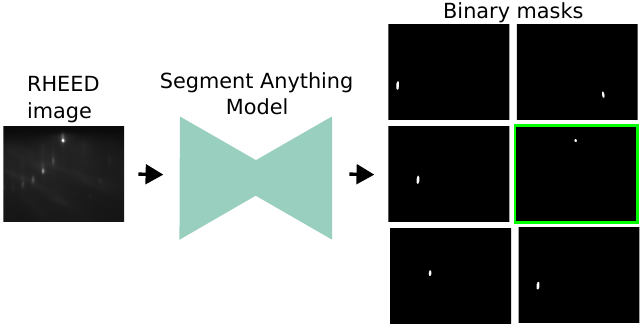}
    \caption{\small Generation of the specular spot mask. Segment Anything Model (SAM) \cite{kirillov_segment_2023} generates a separate mask for any object in the input image. The goal is to manually save a couple of an image and the corresponding specular spot mask (framed in green) in order to build a database for the training of a model that will generate only the specular spot mask.}
    \label{fig:SAM_illustration}
\end{figure}

A dataset of image-mask pairs is set up to train a segmentation model. To this end, binary masks are generated for each spot and line on the RHEED images using the Segment Anything Model (SAM) \cite{kirillov_segment_2023} released by Meta.  SAM operates as a general-purpose segmentation framework, capable of identifying and segmenting objects in basically any image inputs.

To generate a database for training a specular spot detector, we let the SAM generate masks from RHEED images leading to the generation of multiple masks for each image as shown on Figure \ref{fig:SAM_illustration}. The mask corresponding to the specular spot, framed in green on Figure \ref{fig:SAM_illustration}, is manually selected in a total of 1 026 images with $ (512 \times 688) $ dimension. After data-augmentation via shifting of images and masks to four sides and cropping, we obtain a dataset of 5 130 image-mask pairs with $ (290 \times 368) $ dimension. 80 \% of the data is used for training and 20 \% for validation. Subsequently, a test is carried out with 9 632 images which are not among those used for training and test. RHEED videos are captured at 24 frames per second, with wafer rotation at 4 rpm in the MBE chamber of RIBER, Compact 21 DZ (C21DZ).

\subsubsection{Neural network architecture}

\begin{figure}[t]
    \centering
    \includegraphics[width=\linewidth]{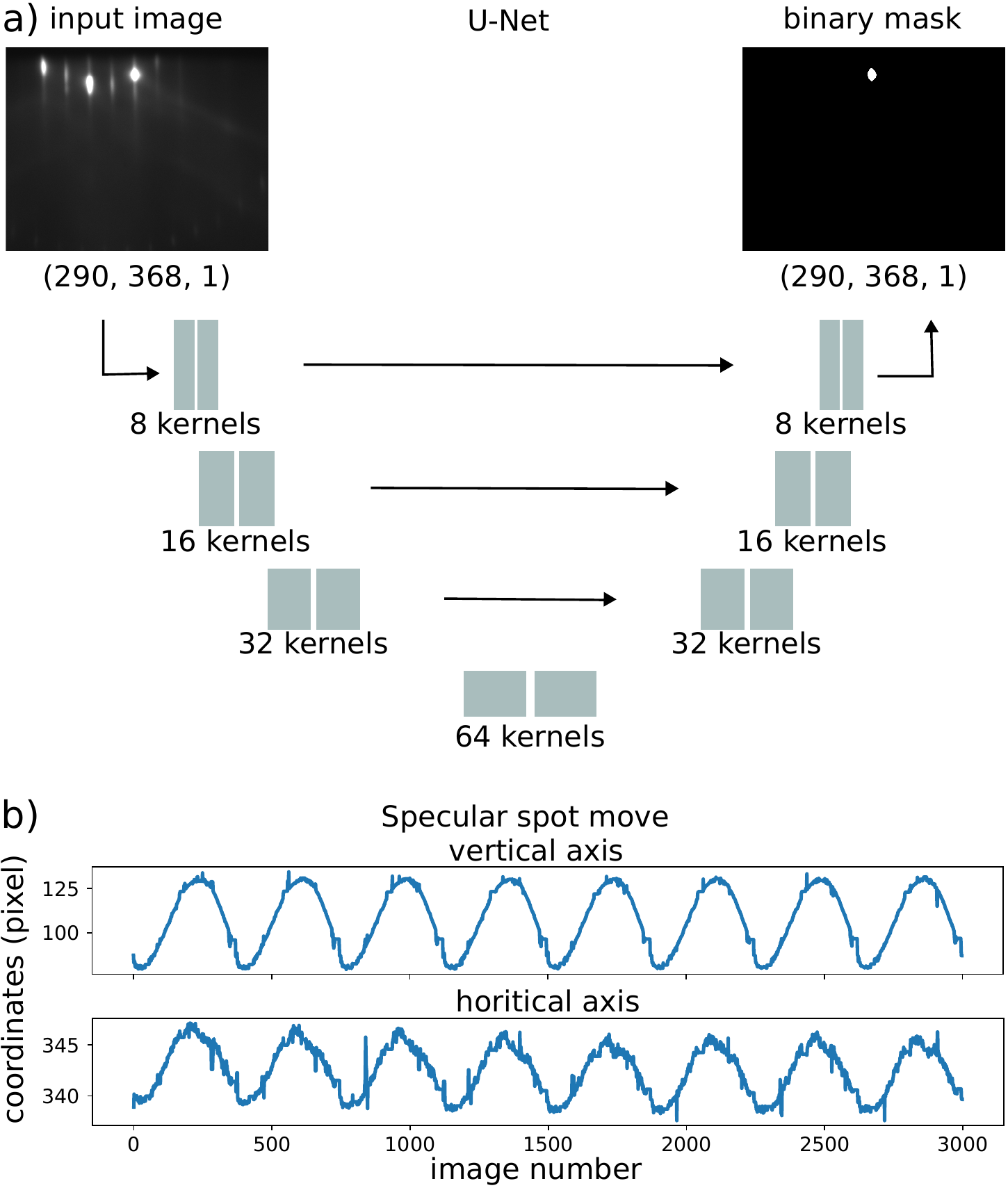}
    \caption{\small a) Sketch of the U-Net model for image segmentation. The model takes a RHEED image as input and reduces the spatial dimensions through an encoder network while expanding the channel dimension. Each rectangular block stands for an Identity Mappings ResNet block \cite{leibe_identity_2016} which is illustrated in Figure \ref{fig:kikuchi_lines_and_model_architecture}d. The decoder, mirror network of the encoder, constructs a matrix with the same spatial dimensions than the input with two channels, each representing the probability of a class (specular or not). The mask is determined by taking the number (1 or 0) of the class with the highest probability value. This leads to a binary mask with ones at the specular spot area and zeros elsewhere. b)  The specular spot coordinates from the first 3k images in the test data. The movement of the specular spot in vertical and horizontal in the test images is determined by the segmentation model via binary masks. When the spot is lost, due either to the electron beam being blocked by the sample holder hooks or to a deterioration in image quality, its last known position is kept.}
    \label{fig:segmantation_model_and_sp_move}
\end{figure}

The model is trained to generate a binary mask for the specular spot from the RHEED image. The U-Net model architecture, depicted in Figure \ref{fig:segmantation_model_and_sp_move}, comprises an encoder and a decoder made up of Identity mapping ResNet blocks \cite{leibe_identity_2016}. The encoder takes an image as input, extracts features through multiple ResNet blocks with increasing number of kernels. The dimensionality reduction is performed using convolution layers with strides as this enables the model to learn the downsampling process. The encoder is followed by a bottleneck of 64 kernels that is fed to a decoder, a mirror network of the encoder with the convolution layers replaced by transpose convolutions with strides for the upsampling. The decoder produces a matrix of two channels, each containing pixel-wise probability for the two classes (specular or not). Thus, the binary mask is determined taking the number of the class with the highest probability. The specular spot coordinates correspond to the center of gravity of the area where the pixels are equal to one. The model is trained with \textit{SparseCategoricalCrossentropy} loss function in keras with \textit{Adam} optimizer using \textit{sigmoid} activation function in the last layer.

\begin{figure*}[t]
    \centering
    \includegraphics[width=0.8\linewidth]{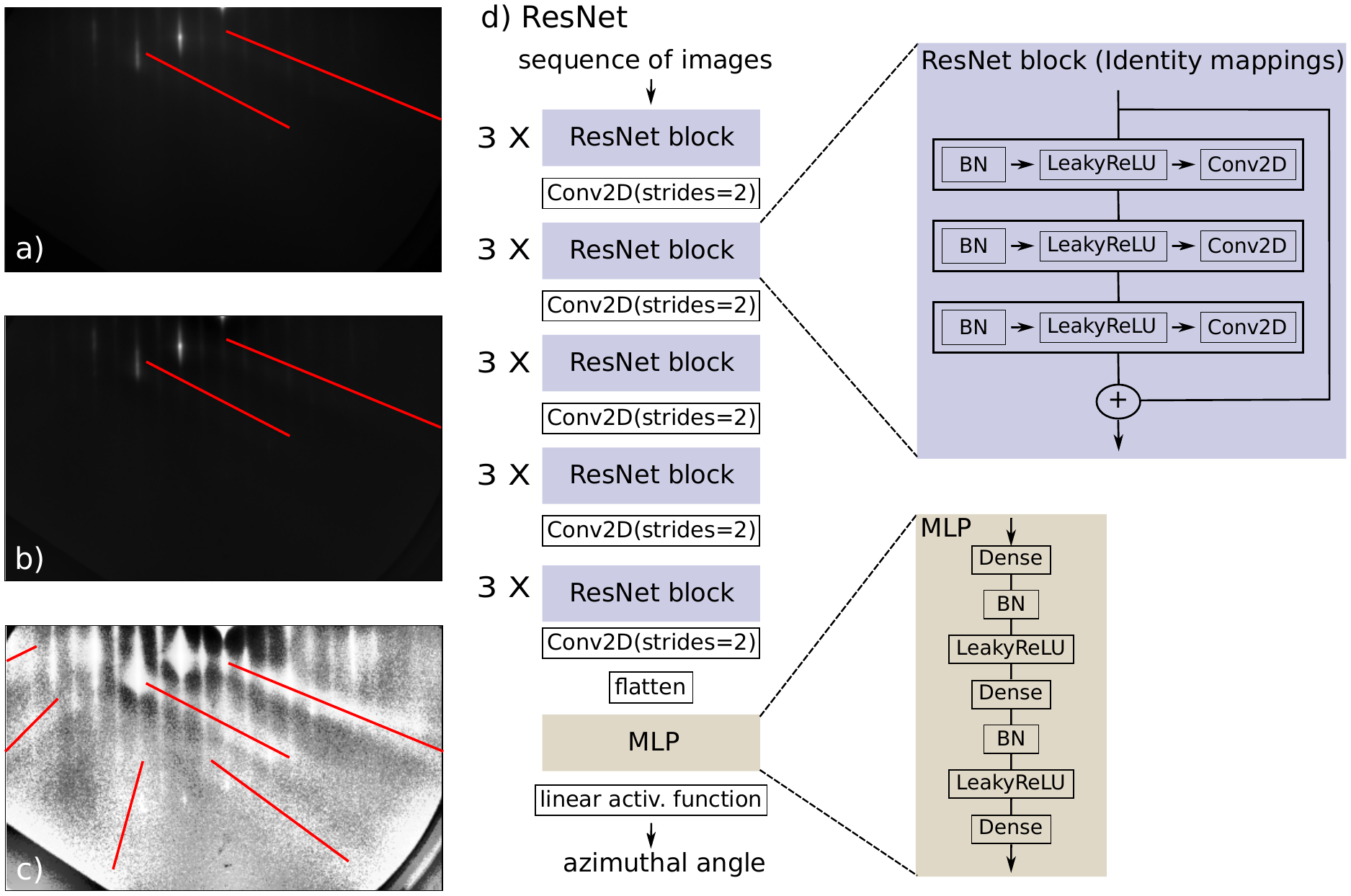}
    \caption{\small l Image preprocessing using Kikuchipy \cite{anes_processing_2020}. Two techniques are applied on the a) original image: first b) removing of dynamic background and then c) adaptive histogram equalization to highlight the Kikuchi lines. The result is shown at each step and the red lines denote the bare-eye visible lines. c) Residual neural network architecture for azimuthal angle. The block of identity mappings ResNet on the top right is called "residual unit" in \cite{leibe_identity_2016} and contains three sub-blocks each consisting of a batch normalization, LeakyReLU activation function and 2D convolution. The result of the last convolution is added to a shortcut of the block input. The architecture of the whole neural network on the left is made of stacked identity mappings ResNet blocks each followed by a single 2D convolution layer for dimensionality reduction to extract informations from the input images. Afterwards, the features go through a \textit{flatten} layer and are fed to an MLP to process the extracted features and predict the azimuthal angle value.}
    \label{fig:kikuchi_lines_and_model_architecture}
\end{figure*}

\subsubsection{Specular spot coordinates}

Once the segmentation model has demonstrated satisfactory performance on the validation data, it is tested on new data to examine its accuracy under real-world conditions. A total of 9 632 images are reserved for this purpose. The movement of the specular spot across the test images is shown in Figure \ref{fig:segmantation_model_and_sp_move}b. The elliptical movement of the specular spot can be seen in these curves, with the vertical displacement having a greater amplitude than the horizontal one. When the spot is lost from view in the black images, its last known position is retained, although this will not allow any useful information to be extracted.

\subsection{Azimuthal angle determination} \label{sec:Azimuthal angle determination}

In addition to the specular position, the ARHEED construction requires an accurate measurement of the angle between electron beam and crystal orientation, the azimuthal angle (the wafer rotation angle). The quality of the Azimuthal RHEED depends on the accuracy with which this angle is determined as well as on the precision of the previously extracted positions. This subsection aims to retrieve that angle from raw RHEED images using a deep learning approach.

\subsubsection{Dataset preprocessing}

As explained before, the sample holder in the ultra high vacuum chamber is magnetically coupled to an external motor, often leading to non-smooth rotation. However, the sample holder has a significant physical weight. The faster we spin it, the less perturbation of the rotation by mechanical play and friction will occur thanks to the inertia of the sample holder. Therefore, angle accuracy is higher under fast spinning, but the ARHEED resolution reduces for high rotation speed. Good angle resolution requires rotation speeds in the order of 3-4 rpm, where friction and mechanical play have a very noticeable impact on the rotation. For this reason, we train a ResNet model for the azimuthal angle regression using a dataset of 26 300 images captured when the sample is rotating at 12 rpm for training. At this speed the rotation is very uniform and the relative angle can be accurately determined even without dedicated angular coder hardware, for example relative to a reference RHEED image.
The dataset is split in such a way that  70 $\%$  is used for the model training, 15 $\%$ for validation and 15 $\%$ for test. For the ARHEED construction, the test dataset of the specular beam tracking in the above section is used. The raw test images have $ (512 \times 688 ) $ dimension in 16-bit (65536 grayscale) and are captured at 4 rpm. In preprocessing, we reduce the image dimensions to  $ (137 \times 229 ) $ and at the same time normalize pixel values between 0 and 1. 
After tests of different size scale factors, we choose the smallest dimensions without noticeable impact on the results. Sequences of 6 images bring better accuracy for the regression task thus the model input dimensions are $ (137 \times 229 \times 6) $, stacking six consecutive RHEED images along the channel dimension. We define the prediction angle to be the angle corresponding to the first image of each sequence.  For a real-time application, the oldest, last image would be removed from the stack and the newest image, obtained from the RHEED camera, would be added as new first image for each prediction. 
We use the open source software Kikuchipy \cite{anes_processing_2020} for RHEED image pre-processing, which allows to increase the contrast of the Kikuchi lines, thereby enabling the model to focus on their movement. We found that this pre-processing significantly improves the training convergence of the azimuthal angle prediction network. Kikuchi lines depend mainly on the bulk of the crystal \cite{pawlak_analysis_2021}, which leads us to believe that this is a robust method for deducing the rotation angle from RHEED images.
Two pre-processing steps are applied on the images using Kikuchipy: removing of dynamic background and adaptive histogram equalization, as illustrated in Figure \ref{fig:kikuchi_lines_and_model_architecture}b and \ref{fig:kikuchi_lines_and_model_architecture}c, respectively. Figure \ref{fig:kikuchi_lines_and_model_architecture}c shows that the contrast of the Kikuchi lines is signficiantly increased after this pre-processing.
Note, that the neural network is trained on the full RHEED images, apart from the Kikuchi lines contrast pre-processing, we implemented no further techniques that would force the models to use exclusively Kikuchi lines for angular regression. To study the impact of Kikuchi visibility or signal-to-noise ration, synthetic data would be required, which would offer the possibility to arbitrarily tune the Kikuchi line contrast or the amount of noise only on the Kikuchi lines.

\subsubsection{Neural network architecture}

An Identity Mappings Residual Network as reported by He Kaiming et al. \cite{leibe_identity_2016} is used for characteristics extraction. Figure \ref{fig:kikuchi_lines_and_model_architecture}d shows the architecture of our RHEED-angle regression model, the numbers on the left of the blocks indicate how many times the residual units are stacked before the single \textit{Conv2D} layer with strides equal to 2 for dimensionality reduction. Our residual neural network architecture consists of residual units involving three convolution layers each preceded by batch normalization and LeakyReLU activation function. A convolution shortcut is used only for dimensionality matching. After following the main residual network, a \textit{flatten} layer combines the extracted features and passes them to a multilayer perceptron consisting of two dense layers each followed by a \textit{Batch Normalization} and a \textit{LeakyReLU} activation function. Finally, the network output is a single linear neuron for the angle output.

\subsubsection{Azimuthal angle prediction}

\begin{figure}[t]
    \centering
    \includegraphics[width=\linewidth]{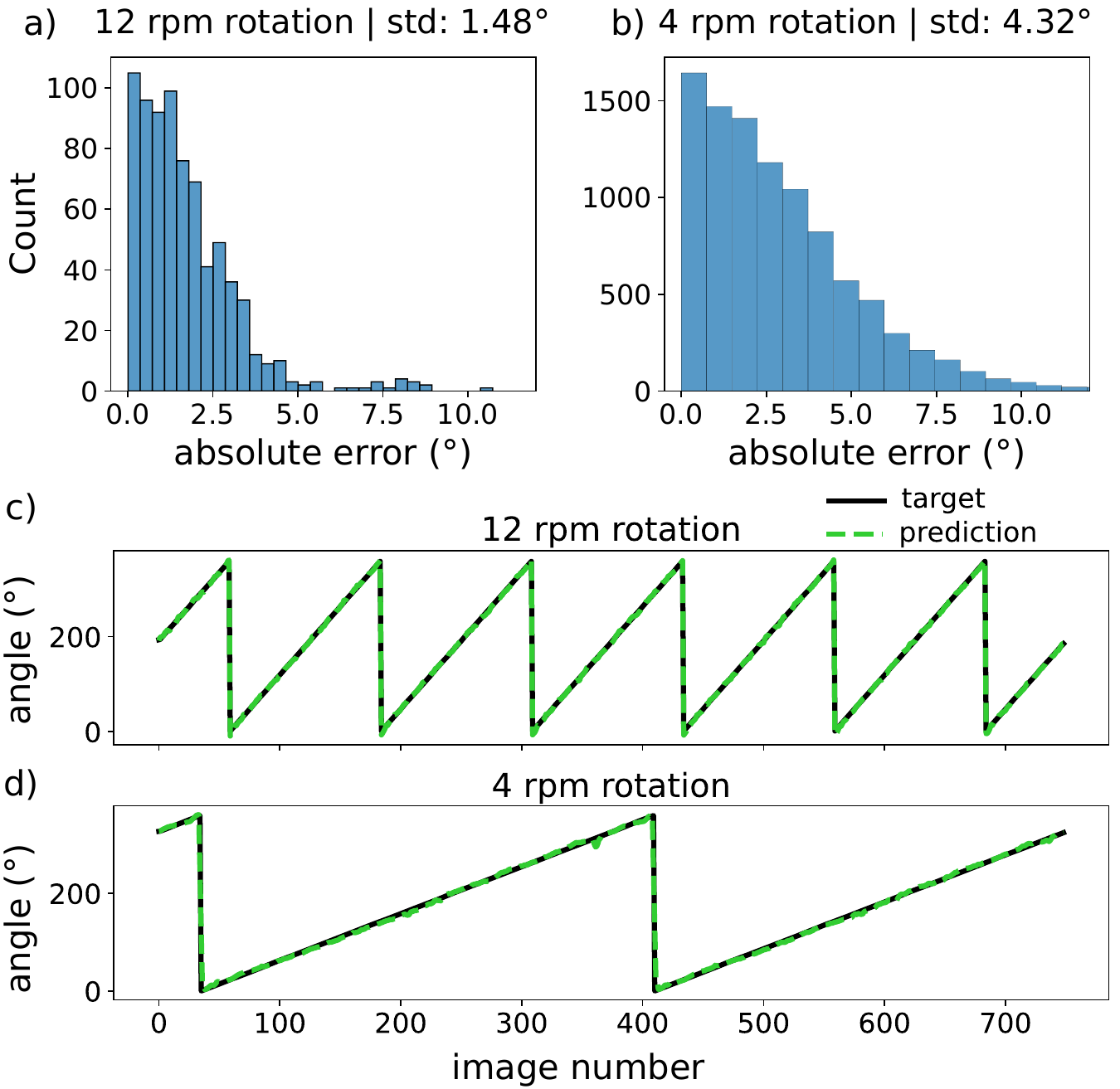}
    \caption{\small Regression model accuracy. The histogram of the absolute error between the prediction and the target values for a) the test data from 12 rpm patterns, where the error is roughly distributed around zero with a standard deviation of 1.48$ ^\circ $ and b) the test data from 4 rpm patterns, where the error is also roughly distributed around zero but with a standard deviation of 4.32$ ^\circ $. The azimuthal angle prediction (green dashed line) versus target values (black line) for RHEED images captured during c) 12 rpm rotating sample and d) 4 rpm rotating sample for a random window of 749 images in each test dataset (12 rpm and 4 rpm).}
    \label{fig:12_vs_4_rpm_predictions}
\end{figure}

Our MBE is equipped with an angle-encoder that gives us a precise measurement of the azimuthal angle, which is not the case for a standard MBE chamber. Thus, we are capable of quantifying the angular prediction error. First we test the model on data from 12 rpm rotating sample.  The histogram of the absolute error is shown in Figure \ref{fig:12_vs_4_rpm_predictions}a), the error is roughly distributed around zero with 1.48$ ^\circ $ standard deviation. In addition, Figure \ref{fig:12_vs_4_rpm_predictions}c) shows the ResNet prediction plot versus the target values, where we can see superimposed curves demonstrating a good prediction accuracy. At this speed, the inertia of the sample holder is large enough to ensure a smooth rotation and we thus have angular values with high confidence.

In a second step, we test the model with data captured during 4 rpm rotating substrate. To emulate the same image sequence characteristics as in the training samples obtained with 12 rpm, we only use every third image for sequence stacking under 4 rpm acquisition. The statistics of the absolute error of the 4 rpm test is shown in Figure \ref{fig:12_vs_4_rpm_predictions}b, showing a distribution around zero with 4.32$ ^\circ $ standard deviation. Figure \ref{fig:12_vs_4_rpm_predictions}d depicts the ResNet predictions as well as the target values. On the predicted angles we apply a Savitzky-Golay filter for smoothing of high-frequency noise. At this speed, the rotation is significantly less smooth due to the mechanical play and friction at the magnetic coupler. Despite these perturbations, the model predicts well the azimuthal angle.

Angular precision of a few degrees is sufficient for the following demonstration, where we distinguish between (2x4) and c(4x4) surface reconstructions. However, to extract more quantitative information such as lattice mismatch, atomic composition or exact phase transition moments, around 4~degrees accuracy may start to be limiting.
As discussed in more detail below, in a production environment, the network should be trained on a broader and larger dataset, also containing different rotation speeds. This should improve
the angular accuracy since larger datasets typically improve generalization (here to other
rotation speeds). One could also imagine a pre-training step on calculated Kikuchi images (see also below). 
Finally, the accuracy of our dataset is limited by the assumption of perfectly homogeneous sample holder rotation, as the angular measurement is performed only once per full rotation. Optimizing the data set carefully for the highest possible angular accuracy may also further improve fidelity.

\subsection{Azimuthal RHEED plotting and interpretation}

\begin{figure*}[t]
    \centering
    \includegraphics[width=0.95\linewidth]{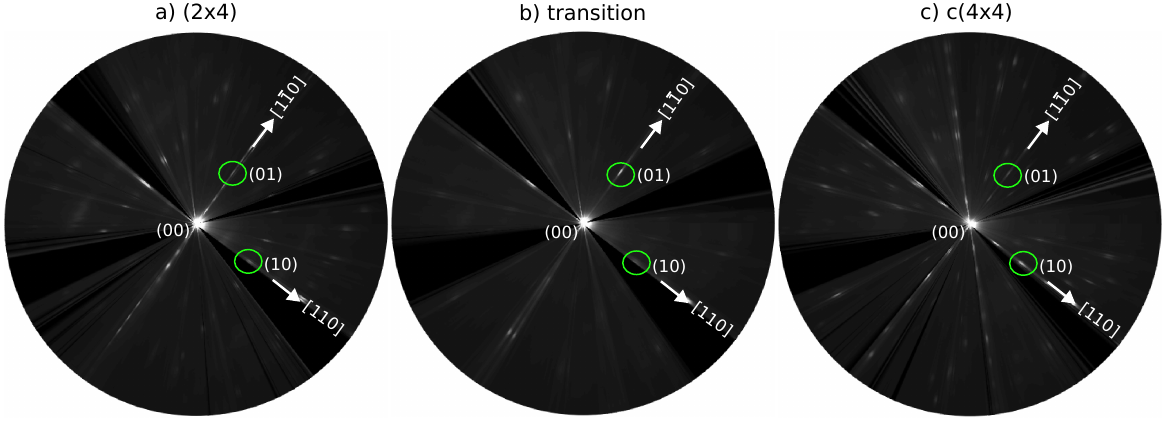}
    \caption{\small Surface reconstruction determination from azimuthal RHEED. Each azimuthal RHEED pattern contains information from two sample rotations. The large black segments correspond to the electron beam being blocked by the hooks of the sample holder. We set a threshold so that these parts appear dark black. This threshold sometimes weeds out other images that are initially almost black everywhere except at the specular point, this corresponds to the dark lines visible at certain angles. These patterns belong to a) $ (2 \times 4) $, b) transition and c) $ c(4 \times 4) $ surface reconstructions. The (00) point with the greatest intensity corresponds to the specular spot while the (01) and (10) spots correspond to the first order diffractions in [1$\overline{1}$0] and [110] directions, respectively.}
    \label{fig:ARHEED_transition}
\end{figure*}

Now that we have established models capable to determine the position of the specular spot and the azimuthal angle, we are now capable to combine these information to create ARHEED images exclusively from the raw RHEED patterns. The Azimuthal RHEED provides access to a wider spectrum of information than a single RHEED image. Each ARHEED pattern in  Figure \ref{fig:ARHEED_transition} is obtained using RHEED patterns corresponding to two sample rotations by averaging the redundant data and interpolation of missing angles. The large black segments correspond to moments during the rotation when the electron beam is blocked by the hooks of the sample holder. The first information that can be extracted from ARHEED is the surface reconstruction which refers to the rearrangement of surface atoms of the material under growth. The $ (001) $ GaAs surface can be terminated by either Ga or As atoms and has two dangling bonds for each surface atom that experiences inter-atomic forces from only the bulk side \cite{ohtake_surface_2008}. Due to this imbalance and in order to eliminate such dangling bonds, the surface atoms undergo complex reconstructions with different spacing and symmetry. These reconstructions are formed critically depending on the preparation conditions \cite{ohtake_surface_2008} such as temperature and affect the surface morphology and electronic properties of the material. Therefore, monitoring surface reconstruction provides valuable information for optimizing growth conditions and controlling material properties.  We continuously created ARHEED representations while decreasing the temperature of a previously-heated As-stabilized GaAs buffer layer. Upon temperature decrease, the surface termination undergoes a transition from $ (2 \times 4) $ to $ c(4 \times 4) $ surface reconstruction. This can be observed in the ARHEED shown in Figure \ref{fig:ARHEED_transition}, where a sample during the $ (2 \times 4) $ reconstruction is shown (left), one at a temperature corresponding to the transition (center) and finally an ARHEED corresponding to a $ c(4 \times 4) $ surface reconstruction (right).

\subsection{Considerations about performance and potential real-time capabilities}

Finally, we want to discuss current shortcomings and potential improvements for our method. For all the results shown here, our method is applied on pre-recorded data. The most interesting application would however be of course in-situ real-time generation of ARHEED plots. 
To assess the real-time potential, we first need to have a look at the inference speed of the neural networks, as well as at the time required for pre- and post-processing.  In inference, the image segmentation takes 39\,ms (image resize + U-Net prediction) for one image, and the regression takes 345\,ms (image resize + kikuchipy: 267\,ms and ResNet 78\,ms) for one image sequence.
Post-processing consists in extracting a slice of the RHEED image starting at the specular spot, inserting this in a numpy array, and updating the angular interpolation. These steps are very fast (faster than 1\,ms) and can thus be ignored in the timing considerations.

In consequence, timing of our workflow for single-image processing is too slow for real-time prediction at a typical 24 fps video acquisition speed. It would as-is suffice only for roughly 3~RHEED images per second. However, there are several straightforward acceleration possibilities to enable real time or quasi real time analysis.
We want to process all captured RHEED images to obtain the best possible angular resolution in ARHEED. However, at typical MBE growth rates, it is not necessary to have response times in the millisecond regime, so we do not need the results displayed in true real-time. 
It would be possible to acquire a certain number of RHEED images in a buffer memory before batch processing all of them in one row. Deep learning models are highly optimized for batch evaluation. Our neural networks process a batch of 24 images in roughly 300\,ms (angle resnet + specular spot segmentation), significantly faster than 24 separate single inferences. Resize + Kikuchipy for a batch of 24 images requires additional 310\,ms, which also is strongly profiting from batch processing. In sum, such batch evaluation would be sufficiently fast for processing of all acquired images in quasi-real time, with an ARHEED refresh rate of $1$\,second (corresponding to the image buffer) and a maximum display delay of roughly $1.6$\,seconds (image buffer + processing time).

The above described quasi-real-time approach would be possible with the tools as described in this work, without any modifications except for the use of batch processing. However, true real-time processing should also be possible through more sophisticated technical performance optimizations, by replacing Kikuchipy with better optimized algorithms, and optimizing the neural network for inference throughput. 
For example, using highly optimized vision models instead of our home-made resnet. The overall architecture could be easily optimized by splitting the pipeline in two stages. A first image-processor network could compress image-per-image information to a latent space. In a second stage, instead of using sequences of full images for angle regression, sequences of latent vectors could be interpreted by a fast and lightweight model. This was demonstrated in an earlier work \cite{khaireh-walieh_monitoring_2023}.

Our test system uses an Intel 12th generation i9 12900 CPU with 128GB of RAM, and an NVIDIA RTX 3090 GPU with 24GB of VRAM. The Nvidia GPU is never utilized by more than roughly 30\,\% of its capacity, indicating that memory transfer overhead is limiting the performance of our small neural networks, and that the models would run well also on less power-hungry hardware. 
In memory requirements, for image pre-processing and Kikuchipy the python memory overhead dominates (roughly 300\,MB of RAM are required for processing a sequence of 24 images). The neural networks (angular regression and specular spot segmentation) require, for a batch of 24 frames, roughly 3GB VRAM each, which is sufficiently frugal to be run on entry level consumer GPUs.

Please note also, that the method as described here works only under the exact conditions that are captured by the training data. Our model does for example not work well on data acquired a few months after the training period, which we attribute to increased deposition on the view-port and variations in the electron beam configuration. We assume that the method can be extended to different growth environments using the fine-tuning technique. This enables a neural network model to adapt on new data that follows similar underlying correlations as the pre-training dataset. Fine-tuning requires less resources than training a model from scratch, yet the data should not be too different. We foresee that fine-tuning should work with similar RHEED systems (same acceleration voltage and similar distances), but full re-training may be necessary if the configurations are too different.
It may also be possible to create a ``foundation model'' from fully synthetic data using computed Kikuchi line figures \cite{pawlakAnalysisKikuchiLines2021}. Such model would not be useful for any actual application, but may be a versatile starting point for fine-tuning training on data from specific growth chambers.

\section{Conclusions}

In conclusion, the presented techniques enable the construction of the Azimuthal RHEED from only raw RHEED patterns. This task requires the specular spot position and the azimuthal angle to be determined with precision. The position is tracked using a segmentation U-Net model trained on a dataset labeled with the help of SAM \cite{kirillov_segment_2023}, a general-purpose segmentation model released by META. 
The determination of the azimuthal angle is done by regression using a ResNet model based on identity mappings \cite{leibe_identity_2016}. The training dataset of the azimuthal angle model is recorded while the sample is rotating at a stable, high speed of 12 rpm. 
Subsequently, it is used to predict angles at low rotation speed (4 rpm), where the angular stability is reduced due to mechanical play and friction at the magnetic coupler to the external motor. 
Each ARHEED pattern is obtained from processing two full rotations through averaging and interpolation of the angles. Furthermore, we demonstrated that the technique is capable to track the transition between different GaAs surface reconstructions during cooling down of the sample. Note that the proposed deep learning models are trained on data from one experiment. 
To work effectively on data from different experiments, the models need to be trained on a variety of data. Future work includes therefore generalization of the deep learning models to multiple materials, not limited to GaAs. 
Initiatives to gather, homogenize and publish experimental RHEED data in open source repositories would be very helpful for further developments.
A possible route for future improvements and generalization could be pre-training of foundation models, using synthetic trajectory data for the specular spot U-Net model, or simulated Kikuchi line images for the angular regression network.
We believe that our method is a highly valuable RHEED analysis technique for production MBEs, which are not equipped with angular in-situ measurement.

\subsection*{Supplementary Material}
A supplementary video of the deep learning constructed ARHEED during the transition between ($2\times 4$) to c($4\times 4$) is available online.

\subsection*{Disclosure}
The authors declare no competing financial interest.

\begin{acknowledgments}
This study benefited from the support of both the LAAS-CNRS micro and nanotechnologies platform, member of the French RENATECH network, and the EPICENTRE common laboratory between Riber and LAAS-CNRS. This work was supported by the French Agence Nationale de la Recherche (ANR) under grant ANR-22-CE42-0021 (project VERDICT).
\end{acknowledgments}

\bibliography{Azimuthal_RHEED_DL.bbl}

\clearpage

\section*{For Table of Contents Use Only}
\textbf{``\doctitle''}

\noindent Abdourahman Khaireh-Walieh, Alexandre Arnoult, Sébastien Plissard, and Peter R. Wiecha

\vspace{1cm}
\paragraph*{Graphical TOC:}
\begin{center}
	\includegraphics[width=\columnwidth]{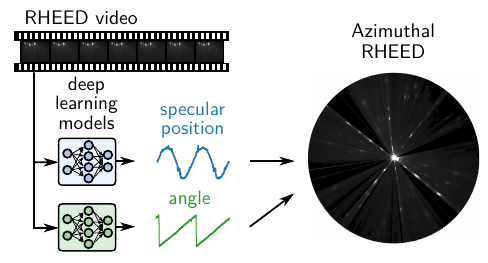}
\end{center}

\vspace{1cm}
\paragraph*{Synopsis:}

Azimuthal RHEED (ARHEED) is important because it enables straightforward interpretation of RHEED patterns for detecting properties such as crystal orientations or periodicities during MBE growth with substrate rotation. Our approach constructs ARHEED using deep learning methods by segmenting the specular spot with a U-Net and by regressing the azimuthal angle with a ResNet. As all information is extracted purely from raw RHEED images, our method does not require specialized hardware for rotation angle surveillance and also works with non-ideal substrate positioning.

\clearpage

\end{document}